\documentstyle[prl,aps,twocolumn,psfig]{revtex}
\begin{document}
\twocolumn[\hsize\textwidth\columnwidth\hsize\csname
  @twocolumnfalse\endcsname
\vspace{-0.5in}
\draft
\title{Implications of the possibility that $\sin{2 \beta}$
is small}
\author{Jo\~ao P.\ Silva\cite{ISEL}}
\address{Stanford Linear Accelerator Center, Stanford University, Stanford,
	 CA 94309, USA}
\author{Lincoln Wolfenstein}
\address{Department of Physics,
Carnegie Mellon University, Pittsburgh, Pennsylvania 15213}
\date{\today}
\maketitle
\begin{abstract}
Recently,
the Babar and Belle collaborations have reported their first measurements
of the CP-violating asymmetry in $B_d \rightarrow \psi K_S$,
and more precise results will follow soon.
We discuss what a future evidence for small $\sin{2 \beta}$
could mean,
contrasting the usual possibility of new physics in the $B_d$ system,
with the interesting alternative that the new physics effects are
confined to the kaon system.
\end{abstract}

] 

The first result to come from $B$-factories is the CP-violating
asymmetry in the decay $B_d \rightarrow \psi K_S$.
This measures $\sin{2 \tilde \beta}$ which,
in the phase convention in which the decay amplitude is real,
coincides with the phase of the $B_d - \overline{B_d}$
mixing matrix element $M_{12}(B_d)$.
In the Standard Model (SM)
$\tilde{\beta} =
\beta =
\arg{\left(- V_{cd} V_{cb}^\ast/ V_{td} V_{tb}^\ast\right)}$.
We will use throughout the Wolfenstein parametrization \cite{Wol}
of the CKM matrix \cite{CKM},
and the corresponding phase convention.
Thus,
$\beta$ is the phase of $V_{td}^\ast$ and
$\sin{2 \beta} = 2\eta(1-\rho)/((1-\rho)^2+\eta^2)$.
The constraints on $\rho$ and $\eta$ from the parameter
$\epsilon_K$ in kaon decays,
combined with those from $x_s$ and $|V_{ub}/V_{cb}|$,
yield $16^\circ \leq \beta \leq 34^\circ$ and 
$38^\circ \leq \gamma \leq 81^\circ$ at the 95\% CL \cite{AliLon},
corresponding to a correlated region bounded roughly by 
$0.24 \leq \eta \leq 0.50$ and
$0.07 \leq \rho \leq 0.38$ \cite{AliLon}.
Here $\gamma= \arg{\left(- V_{ud} V_{ub}^\ast/ V_{cd} V_{cb}^\ast\right)}$,
and $\tan \gamma = \eta/\rho$.

Initial measurements by the CDF collaboration \cite{CDF} found
$\sin{2 \tilde \beta} = 0.79^{+0.41}_{-0.44}$.
Recently the Babar \cite{Babar} and Belle \cite{Belle}
collaborations announced their results to be
$\sin{2 \tilde \beta} = 0.12 \pm 0.37 ({\rm stat}) \pm 0.09 ({\rm sys})$
and $\sin{2 \tilde \beta} = 
0.45^{+0.43}_{-0.44} ({\rm stat})^{\, +0.07}_{\, -0.09} ({\rm sys})$
(combining with $J/\psi K_L$),
respectively.
Although the errors are large,
and a detailed numerical analysis is still unwarranted,
it is interesting to study the possibility that
$\sin{2 \tilde \beta}$ is significantly lower than the value
allowed by the SM.
This is what we consider here.
As an example,
we look at the consequence of a value
$\sin{2 \tilde \beta} = 0.2$,
which requires new physics beyond the Standard Model.

We contrast two possibilities.
The first possibility is that all the new physics is in the kaon system;
since CP violation is very small for kaons in the
Standard Model,
there is great sensitivity to new physics.
The second possibility is that there is new physics in
$B_d - \overline{B_d}$ mixing;
in the Standard Model this mixing is proportional to 
$A^2 \lambda^6 \eta = 10^{-5} - 10^{-4}$ and,
so,
new physics effects could be significant.

Consider the possibility that the new physics is confined to
$K^0 -\overline{K^0}$ mixing.
This is the original evidence for CP violation 
and is used in determining a lower limit on $\eta$,
corresponding to a lower limit on $\beta$.
It is interesting to note that there are no current experiments on 
kaons that can directly detect such a new physics contribution.
This may be possible with $K_L \rightarrow \pi^0 \nu \bar \nu$
\cite{GroNir},
but the remarkable feature we stress here is that such effects
might be detected first in $B$ decays.
As was noted long ago,
the CP-violating part of the 
$K^0 -\overline{K^0}$ mixing matrix can be explained by a superweak 
interaction with an effective coupling as small as 
$10^{-10} G_F$ to $10^{-11} G_F$ \cite{superweak}.
Assuming no other new physics,
then $\tilde \beta = \beta = 0.5 \arcsin{0.2}$,
and the measurement of $|V_{ub}/V_{cb}|=0.093 \pm 0.014$
implies $0.27 \leq \sqrt{\rho^2 + \eta^2} \leq 0.58$
at $95\%$ CL \cite{AliLon}.
Intercepting these constraints leads to 
$\gamma = \arctan{\eta/\rho}$ between $4^\circ$ and $16^\circ$. 
(We have used $x_s$ to eliminate the solutions with
negative $\rho$, corresponding to very large values for $\gamma$.)

An argument against this interpretation is the relatively large
value of $\epsilon^\prime_K/\epsilon_K$ observed in the kaon system
\cite{epsilonprime}.
While there are large uncertainties in the Standard Model calculation
\cite{RMP},
even the largest theoretical estimates of the error bars in the hadronic
matrix elements require values of $\eta$ greater than $0.1$ to fit the
recent experimental results.
Thus,
this would seem to require new physics in the CP-violating kaon
decay amplitude as well as in the mixing.
This possibility has been sugested in some models \cite{Masiero}.

An alternative is to consider new physics in the
$B_d - \overline{B_d}$ mixing \cite{Bdmixing}. 
If this is the only source of new physics,
a value of $\eta$ larger than about $0.25$ is required by the
observables in the kaon system.
Conversely,
given the constraint on $|V_{ub}/V_{cb}|$,
once we allow $\eta$ greater than $0.25$,
the question of whether there is new physics in
$K^0 -\overline{K^0}$ mixing is unimportant and
we do not consider it further.
Henceforth,
we will take $\eta \geq 0.25$.

A natural possibility is to assume that there is no significant new
physics in $B_s - \overline{B_s}$ mixing,
on the grounds that the magnitude of $M_{12}(B_s)$ is much larger than
for $B_d$ and, so,
less susceptible to new physics. 
Then,
once $x_s$ is known,
we can extract $f^2_{Bs} B_{Bs}$,
to be combined with the reliable lattice-QCD estimate of
$K=f^2_{Bs} B_{Bs}/f^2_{Bd} B_{Bd} = 1.30 \pm 0.14$ \cite{lattice}
in order to extract $f^2_{Bd} B_{Bd}$.
We may now determine $M_{12}^{SM}(B_d)$ for any value of $(\rho, \eta)$.
Indeed,
it is easy to show that,
in this scenario,
\begin{equation}
M_{12}^{SM} (B_d)
= \frac{x_s \Gamma_{Bs}}{2}
K^{-1}\, \lambda^2 R_t^2\, e^{2 i \beta},
\end{equation}
with $R_t=\sin{\gamma}/\sin{(\gamma+\beta)} = \sqrt{(1-\rho)^2 + \eta^2}$.
Using,
\begin{equation}
M_{12}(B_d)= \frac{x_d \Gamma_{Bd}}{2}\, e^{2 i \tilde \beta},
\end{equation}
we may now determine
\begin{equation}
M_{12}^{\rm new}(B_d) = M_{12}(B_d) - M_{12}^{SM}(B_d),
\end{equation}
for chosen values of $\beta$ and $\gamma$
({\it i.e.}, of $\rho$ and $\eta$).
This is illustrated in figure~1,
where we have used $x_d=0.723$,
$x_s = 20$,
$K=1.3$,
$\lambda=0.22$,
$\sin{2 \tilde \beta}= 0.2$,
$\eta \geq 0.25$,
and taken 
$R_b=\sin{\beta}/\sin{(\gamma+\beta)} = \sqrt{\rho^2 + \eta^2}$
to equal $0.27$, $0.4$, and $0.58$.
In figure~1,
we have also used $\Gamma_{Bs} \sim \Gamma_{Bd}$ and 
divided all vectors by $|M_{12}(B_d)|$.
\begin{figure}
\centerline{\psfig{figure=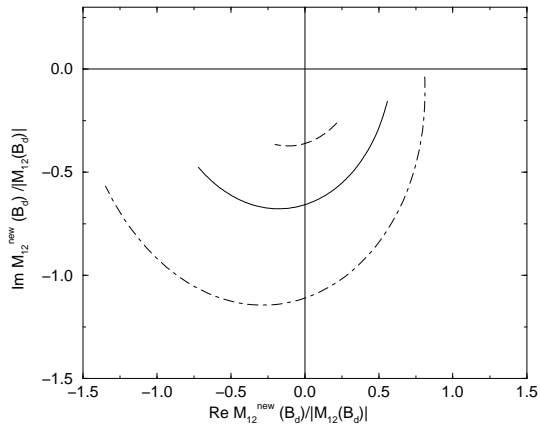,height=2.6in}}
\caption{Position of the tip of the vector $M_{12}^{\rm new}(B_d)$
in the complex plane,
in the usual phase convention,
and in units of $|M_{12}(B_d)|$.
The position of the tip varies as we vary $\rho$ and $\eta$.
The input parameters are
$x_d=0.723$,
$x_s = 20$,
$K=1.3$,
$\lambda=0.22$,
and
$\sin{2 \tilde \beta}= 0.2$.
The dashed,
solid,
and dot-dashed lines correspond to
$R_b = 0.27$, 
$R_b = 0.4$,
and $R_b = 0.58$,
respectively.
\label{fig:1}}
\end{figure}
Figure~2 shows the magnitude of $M_{12}^{\rm new}(B_d)$ as a function
of $\gamma$,
with the same parameters and conventions used in figure~1.
Using the requirement that $\eta \geq 0.25$,
we find that $25^\circ \leq \gamma \leq 155^\circ$ for
$R_b=0.58$,
$39^\circ \leq \gamma \leq 141^\circ$ for
$R_b=0.40$,
and 
$68^\circ \leq \gamma \leq 112^\circ$ for
$R_b=0.27$.
\begin{figure}
\centerline{\psfig{figure=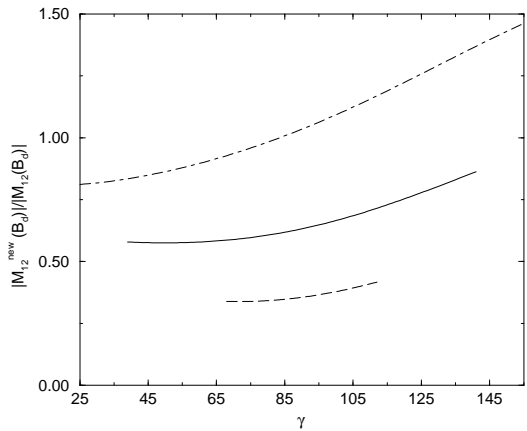,height=2.6in}}
\caption{Magnitude of the vector $M_{12}^{\rm new}(B_d)$,
in units of $|M_{12}(B_d)|$
as a function of $\gamma$.
The input parameters coincide with those in figure~1.
The dashed,
solid,
and dot-dashed lines correspond to
$R_b = 0.27$, 
$R_b = 0.4$,
and $R_b = 0.58$,
respectively.
\label{fig:2}}
\end{figure}

Another possibility is that there is new physics also in
$B_s - \overline{B_s}$ mixing.
Given the current constraints on the $(\rho,\eta)$ plane from
$\epsilon_K$ and $|V_{ub}/V_{cb}|$,
and accepting the lattice estimate for $f^2_{Bs} B_{Bs}$,
there is an upper bound on $x_s$.
If the measurement of $x_s$ were to exceed this upper bound,
one would have an indication of new physics in the $B_s$ system.
Unfortunately,
the lattice prediction for  $f^2_{Bs} B_{Bs}$ is not
as reliable as that for $K$.
Therefore,
only values for $x_s$ considerably above the current upper limit
could be considered as compelling evidence for new physics.
On the other hand,
in the Standard Model and with the usual phase convention,
the phase of $M_{12}(B_s)$ is given by
$\epsilon = \arg{\left(- V_{cb} V_{cs}^\ast/ V_{tb} V_{ts}^\ast\right)}$,
which is of order $\lambda^2 \eta$.
Since this phase is small,
it is conceivable that the new physics might change this phase into
$\tilde \epsilon$,
without significantly affecting the magnitude of $M_{12}(B_s)$.
Large values for $\tilde \epsilon$ would be detected through
the CP-violating asymmetry in $B_s \rightarrow D_s^+ D_s^-$
\cite{NirSil}
or in $B_s \rightarrow J/\psi \phi$
(with a suitable analysis of the angular distributions).

We now discuss how to contrast the two possibilities:
the possibility that all the new physics is in the kaon system;
with the possibility that all the new physics is in the $B$ system.
The most striking difference between the two is in the value
of $\gamma$.
The phase $\gamma$ is less than $16^\circ$ in the first case and
larger than $25^\circ$ in the the second case,
with the exact range depending on the value of $R_b$.
A number of experiments have been discussed in the 
literature to constrain $\gamma$:
i) experiments sensitive to $\sin^2{\gamma}$ \cite{sin2gamma};
ii) the CP-violating asymmetry in $B_d \rightarrow \pi \pi$,
which, once the penguin effect is taken into account,
measures $\sin{(2 \tilde \beta+ 2 \gamma)}$ \cite{pi pi},
iii) the Dalitz plot analysis of $B_d \rightarrow \rho \pi$,
which determines $\sin{(2 \tilde \beta+ 2 \gamma)}$
and also $\cos{(2 \tilde \beta+ 2 \gamma)}$
\cite{rhopi};
iv) and experiments sensitive $\sin{(2 \tilde \beta + \gamma)}$
\cite{2beta+gamma}.
Although some of the relevant channels require some knowledge about
the strong phases,
while others require large statistics,
the quest for $\gamma$ may ultimately allow us to 
distinguish between the two alternatives.

A further very interesting feature is that,
in the second case,
$\beta$ and $\tilde \beta$ are different.
Therefore,
the phase of the penguin amplitude $\beta$ is not equal to the mixing
phase $\tilde \beta$.
This has implications for the analysis of the penguin effects.
For example,
if the penguin effects in $B_d \rightarrow \pi \pi$ were small,
the asymmetry would be given by $\sin(2 \tilde \beta + 2 \gamma)$,
with a leading correction of $2 r \cos(2 \tilde \beta + 2 \gamma)
\sin(\beta + \gamma) \cos(\Delta)$,
where $r$ is the ratio of tree to penguin amplitudes and
$\Delta$ the relative strong phase.
Notice that both $\tilde \beta$ and $\beta$ are involved.

Another interesting signal could come from a pure penguin decay
into a CP eigenstate, such as $B_d \rightarrow K_S K_S$.
If this were dominated by the penguin with the intermediate top quark,
the asymmetry would be proportional to $\sin{(2 \beta-2 \tilde \beta)}$.
Unfortunately,
as pointed out by Fleisher \cite{Flei94c},
this method is obscured by the presence of the penguin diagram
with intermediate charm,
which carries a different weak phase.

Of course,
there is the logical possibility that there is new physics in both the
kaon and $B$ sectors.
The new physics could even account for all the observable
CP-violating effects.
Since $\epsilon^\prime_K/\epsilon_K$ is so small,
such a theory would be superweak-like and, thus,
ruled out by measuring a nonzero value for $\gamma$.

Finally, we comment on the discrete ambiguities \cite{discrete}.
If the measurement of $\sin{2 \tilde \beta}$ in $B_d \rightarrow \psi K_s$
turns out to lie within the range allowed by the Standard Model,
then the discrete ambiguities might be crucial in unearthing the
new physics contributions.
However,
if,
as proposed here,
the measurement of $\sin{2 \tilde \beta}$ yields $0.2$,
one finds a natural explanation in a contribution
from new physics driven by
$2 \beta - 2 \tilde \beta$ of order $20^\circ$.
In contrast,
the discretely-ambiguous possibilities 
$\tilde \beta = \{ 84.2^\circ, 185.8^\circ, 264.2^\circ\}$
would correspond to much larger new physics contributions.

In conclusion,
a small value of $\sin{2 \tilde \beta}$ would be a sign of physics
beyond the Standard Model.
This new physics could be confined to the kaon system 
or be primarily in the $B$ system.
Only a variety of further experiments could distinguish these
two possibilities.

\acknowledgments

We thank Yossi Nir for useful discussions.
This work is supported in part by the Department of Energy 
under contracts DE-AC03-76SF00515 and DE-FG02-91-ER-40682.
The work of J.\ P.\ S.\ is supported in part by Fulbright,
Instituto Cam\~oes, and by the Portuguese FCT, under grant
PRAXIS XXI/BPD/20129/99	and contract CERN/S/FIS/1214/98.

%

%
%
%
%
%
%
%
%
%
%
%
%
%
%

\end{document}